\documentclass[sigconf]{acmart-me}

\usepackage{booktabs} 
\usepackage{url}
\usepackage{color}
\usepackage{enumitem}
\usepackage{multirow}
\usepackage{subfloat}
\usepackage{subfig}
\usepackage{balance}
\setcopyright{rightsretained}
\usepackage[printonlyused]{acronym}

\acmDOI{}

\acmISBN{}

\acmConference[MediaEval'19]{Multimedia Evaluation Workshop}{27-29 October 2019}{Sophia Antipolis, France \\ Github: https://github.com/vlbthambawita/MedicoTask\_2019\_paper\_1} 
\acmYear{}
\copyrightyear{}

\acmPrice{}

\newacro{MLP}  [MLP]  {multi-layer perceptron}
\newacro{MSE} [MSE] {mean square error}
\newacro{MAE} [MAE] {mean absolute error}

\begin{document}
\title{Stacked dense optical flows and dropout layers to predict \\sperm motility and morphology}

\author{Vajira Thambawita\textsuperscript{1,2}, P{\aa}l Halvorsen\textsuperscript{1,2}, Hugo Hammer\textsuperscript{1,2}, Michael Riegler\textsuperscript{1,3}, Trine B. Haugen\textsuperscript{2}}
\affiliation{
\textsuperscript{1}SimulaMet, Norway \ \ \ {}
\textsuperscript{2}Oslo Metropolitan University, Norway\ \ \ \
\textsuperscript{3}Kristiania University College, Norway \\
}
\email{Contact: vajira@simula.no}

%
%
%
%
%

\renewcommand{\shortauthors}{Thambawita et al.}
\renewcommand{\shorttitle}{2019 Medico Medical Multimedia}

\begin{abstract}
In this paper, we analyse two deep learning methods to predict sperm motility and sperm morphology from sperm videos. We use two different inputs: stacked pure frames of videos and dense optical flows of video frames. To solve this regression task of predicting motility and morphology, stacked dense optical flows and extracted original frames from sperm videos were used with the modified state of the art convolution neural networks. For modifications of the selected models, we have introduced an additional multi-layer perceptron to overcome the problem of over-fitting. The method which had an additional multi-layer perceptron with dropout layers, shows the best results when the inputs consist of both dense optical flows and an original frame of videos.
\end{abstract}

%
%
%
%
%

\maketitle

\section{Introduction}
\label{sec:intro}
Our main goal of this task is to predict the sperm motility and sperm morphology from videos of sperm samples. In the 2019 Medico task \cite{medico2019overview}, a video dataset was provided with ground truth values of sperm motility such as progressive motility, non-progressive motility, and immotility, and sperm morphology such as head defects, tail defects, and midpiece and neck defects. This task was introduced as completely new this year, and therefore, we could not find any previous work in previous mediaeval Medico task competitions \cite{pogorelov2018medico, riegler2017multimedia}. In this competition, the VISEM dataset \cite{visem} which contains sperm videos recorded from 85 participants is used.  In the dataset paper, the authors presented baseline mean absolute error values for motility and morphology. Moreover, the importance of computer-aided sperm analysis can be identified from the research works which have been done to develop automatic sperm analysis method in last few decades \cite{mortimer2015future, Urbano2017, Dewan2018}.

Video analysis is a hot research topic in the field of deep learning. Some researchers are experimenting with video classification \cite{video_classifcation}, detection \cite{bovik2010handbook}, segmentation \cite{hampapur1994digital}, and generations \cite{li2018video, tulyakov2018mocogan} for various type of video datasets.  Yue-Hei Ng et al. \cite{video_classfication} experimented with video classification problem using well knows datasets such as sports-1M \cite{data_2014large} and UCF101 \cite{data_ucf101}. In these experiments, they have generated dense optical flow images and row frames of videos to classify 120 seconds long videos. In this paper, we use very short video segments such as nine frames compared to these long segments such as 120s X 30 frames/s. 

To solve this new regression problem of predicting morphology and motility from videos of sperm samples, this paper presents two deep learning methods where we used extracted dense optical flows and raw frames from the videos. In Section \ref{sec:approach}, we are going to present our two types of input data and two types of methods used in our experiments. Then, the results collected from these experiments will be discussed in Section \ref{sec:results_and_analysis}. Finally, the paper ends up with conclusions and future work in Section \ref{sec:conclution}.   


\section{Approach}
\label{sec:approach}
We have selected the pre-trained ResNet-34 \cite{resnet_34} to do some basic experiments of predicting sperm motility and sperm morphology using stacked normal raw video frames and a combination of stacked dense optical flows and raw frames of videos. In this paper, we obtain experimental results using two different types of inputs and from two different types of models.

\subsection{Preprocessing data}
\label{sec:input_data}

\begin{figure}[t]
    \centering
    
    \subfloat[][\centering Original frame]{
        \includegraphics[width=2.4cm, height=2.4cm]{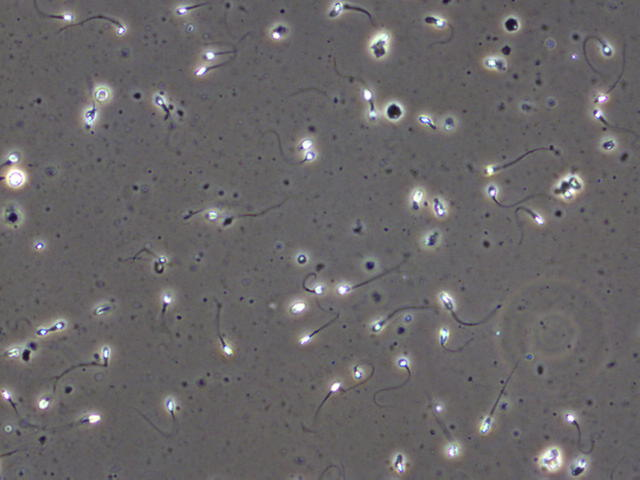}
        \label{fig:original_frame}
    }
    \hspace{2mm}
    \subfloat[][\centering Dense optical flow - stride 1]{
        \includegraphics[width=2.4cm, height=2.4cm]{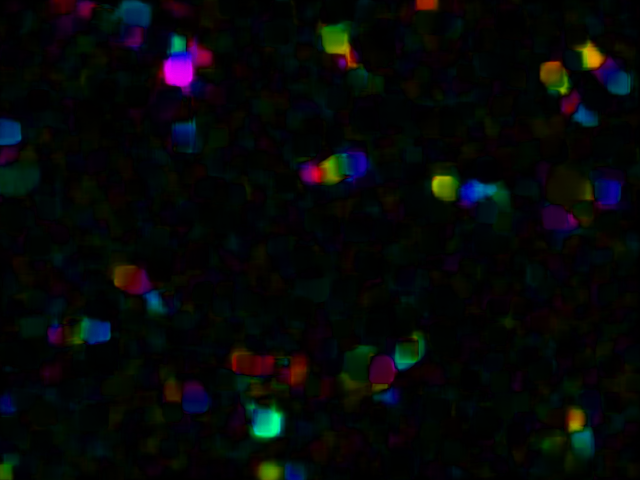}
        \label{fig:stride_1_dense}
    }
    \hspace{2mm}
    \subfloat[][\centering Dense optical flow - stride 10]{
        \includegraphics[width=2.4cm, height=2.4cm]{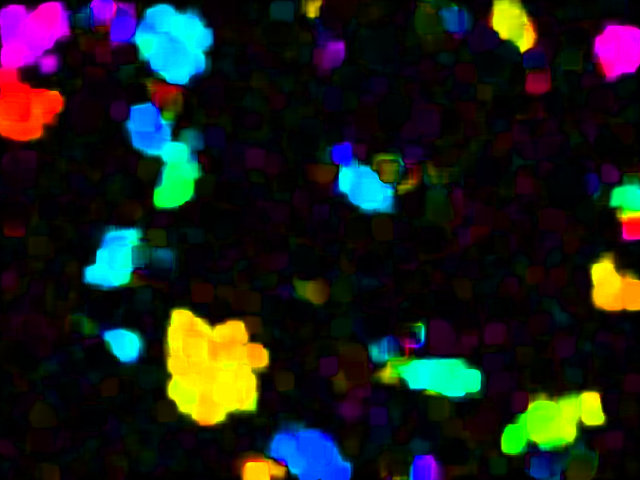}
        \label{fig:stride_10_dense}
    }
    \vspace{-10pt}
    \caption{Sample images used to construct input image stacks into the models}
    \label{fig:my_label}
    \vspace{-10pt}
\end{figure}{}

To find estimates for the sperm motility and sperm morphology, we first preprocessed the input videos to generate two types of input. In the first type \textbf{(dataset - D1)}, we stacked nine consecutive frames from a video to make a single input data point. A sample of a raw frame of a video is given in Figure \ref{fig:original_frame}. Before stacking raw video frames, we converted the RGB format frames of the video into grayscale images and resized them into 256x256. These nine frames represent nine different consecutive frames of a video. Moreover, we collected 250 stacked data points (chunks) from 250 locations in time from a video as described above. 

For the second type of input \textbf{(dataset - D2)},  we generated a tensor with nine channels, which consists of a three-channels (RGB) original video frame (Figure \ref{fig:original_frame}), a three-channels dense optical flow image of stride 1 (Figure \ref{fig:stride_1_dense}), and  a three-channels image of dense optical flow of stride 10 (Figure \ref{fig:stride_10_dense}). The dense optical flow image of stride 1 was generated from two consecutive video frames from a selected location of a video. Then, we generated the stride-10 dense optical flow image using two frames; the first frame of the video chuck and the $10^{th}$ frame of a selected video chunk.  To generate dense optical flows \cite{denseopticalflow} of two different frames of a video, the OpenCV library \cite{opencv} was used with its inbuilt functions.

For both input types, we split the datasets into three folds based on the folds given in the video dataset provided by organizers. Then, a three-fold cross-validation was performed to evaluate our deep learning models which will be introduced in the later sections.


\subsection{Deep learning model implementation}

\begin{figure}
    \centering
    \includegraphics[scale=0.85]{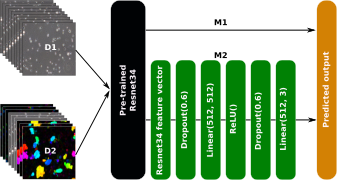}
    \caption{Big picture of our deep learning model: M1 - the base model of Resnet-34 with a three output last layer, M2 - the modified version of Resnet-34 with an additional MLP,  D1 and D2 represent the two different types of input used in our experiments.}
    
    \label{fig:model_architecture}
    \vspace{-10pt}
\end{figure}

For implementation of our deep learning models, we selected Resnet-34 which is larger than the smallest, Resnet-18, and smaller than other large scales Resnet models like Resnet-50, Resnet-101, and Resnet-152. The selections of this intermediate Resnet-34 was done based on expandability of the model by adding additional \ac{MLP} within the available hardware resources (considering memory limitations of the available graphics processing units).  In addition to that,  the pre-experiments were done to identify over-fitting problems of strong models for simpler predictions and  computation time required to finish training. Furthermore, expandability of the number of input channels of the model within the available GPU memory was examined. 

For \textbf{method 1 (M1)}, we modified the input layer of the selected pre-trained Resnet-34 to take nine channel inputs and modified the last layer of the model to output only three values which are representing either three values of sperm motility or three values of sperm morphology.  We used this method as our base model with the two different datasets (D1 and D2) as introduced in Section \ref{sec:input_data} and recorded results collected from this experiment in D1-M1 and D2-M1 rows in Table \ref{tbl:results}. 


In \textbf{method 2 (M2)}, to avoid over-fitting problems of this task, we have embedded additional \ac{MLP} to the end of the network with dropout layers \cite{dropout}. The full structure of this additional MLP is depicted in Figure \ref{fig:model_architecture} using a green colour. The dropout values of this \ac{MLP} were selected using pre-experiments, and it is a hyper-parameter for this model.  The collected results of this method are tabulated in rows D1-M2 and D2-M2 of Table \ref{tbl:results}. 
  
 In the training process of all the above methods, the Adam optimizer \cite{adam} with a learning rate 0.001 was used.  The \ac{MSE} was used as the loss function for back-propagating error, and \ac{MAE} was used for calculating the actual loss of predictions based on ground truth values of motility and morphology.


\section{Results and Analysis}
\label{sec:results_and_analysis}

According to the average \ac{MAE} values shown in Table \ref{tbl:results}, the M2 method with the input type 2 (D2) shows best results among other methods and other input types. This method shows the best \ac{MAE} value of 8.825 for the sperm motility and 5.293 for the sperm morphology.  This improvement of error values can be seen as results of accumulated benefits of showing pre-processed temporal information such as dense optical flows to the model and the additional \ac{MLP} to overcome the problem of over-fitting. Moreover, the added \ac{MLP} in M2 gives better results with both input types (D1 and D2) for both predictions: sperm motility and sperm morphology. We achieved this performance as a result of the pre-processed input data with dense optical flows and the \ac{MLP} introduced to overcome the over-fitting problem.   


\begin{table}[]
\caption{\ac{MAE} values collected from the proposed methods: D1-stacked gray-scale nine consecutive frames, D2-stacked an original frame + a dense optical flow image from two consecutive frames + a dense optical flow from two frames with stride=10; M1 - the basic model of Resnet-34 with modifications of number of input channels and  outputs, M2 - the modified model with an additional MLP with dropout layers}

\vspace{-10pt}
\begin{tabular}{llllcll}
\toprule
&  &  & \multicolumn{2}{c}{Motility} & \multicolumn{2}{c}{Morphology} \\ \cmidrule{4-7}
Input & Method & Fold & \multicolumn{1}{c}{MAE} & Average & \multicolumn{1}{c}{MAE} & \multicolumn{1}{c}{Average} \\ \midrule

\multirow{6}{*}{D1} &
\multirow{3}{*}{M1} 
  & Fold 1 & 9.562 & \multicolumn{1}{l}{\multirow{3}{*}{9.200}} & 5.626 & \multirow{3}{*}{5.649} \\
&  & Fold 2 & 8.959 &                       \multicolumn{1}{l}{} & 5.749 &  \\
&  & Fold 3 & 9.079 &                       \multicolumn{1}{l}{} & 5.573 &  \\ \cmidrule{2-7}
 
 &
\multirow{3}{*}{M2} 
  & Fold 1 & 9.585 & \multicolumn{1}{l}{\multirow{3}{*}{9.185}} & 5.424 & \multirow{3}{*}{5.394} \\
&  & Fold 2 & 9.28 &                       \multicolumn{1}{l}{} & 5.382 &  \\
&  & Fold 3 & 8.689 &                       \multicolumn{1}{l}{} & 5.375 &  \\ \midrule
 
\multirow{6}{*}{D2} &
\multirow{3}{*}{M1} 
  & Fold 1 & 9.044 & \multicolumn{1}{l}{\multirow{3}{*}{9.372}} & 5.933 & \multirow{3}{*}{5.525} \\
&  & Fold 2 & 8.062 &                       \multicolumn{1}{l}{} & 5.394 &  \\
&  & Fold 3 & 11.01 &                       \multicolumn{1}{l}{} & 5.248 &  \\ \cmidrule{2-7}
 
 &
\multirow{3}{*}{M2} 
  & Fold 1 & 8.612 & \multicolumn{1}{l}{\multirow{3}{*}{\textbf{8.825}}} & 5.549 & \multirow{3}{*}{\textbf{5.293}} \\
&  & Fold 2 & 7.873 &  &                    5.463 &  \\
&  & Fold 3 & 9.991 &  &                    4.868 &  \\ 
 
 \bottomrule
\end{tabular}
\label{tbl:results}
\vspace{-5pt}
\end{table}


%
%

\section{Conclusion and Future work}
\label{sec:conclution}
The input with a raw frame and dense optical flows of two difference  stride values show better results compared to the stacked normal frames of videos. Moreover, the modified Resnet-34 model with an \ac{MLP} which consists of dropout layers with high probabilities did achieve better results than the base model in the both cases because it helped to overcome the problem of over-fitting in the training stage. Finally, the combination of the input with dense optical flows and the modified Resnet-34 with an additional \ac{MLP} shows the best overall performance. 

In future work, it is worth to try CNN models with long short-term memory units to capture temporal features of video frames. Moreover, a 3D CNN can be a promising approach for this kind of task because 3D CNN models have capabilities to capture temporal information of videos.

\newpage
\bibliographystyle{ACM-Reference-Format}
\def\bibfont{\small} 
\balance
\bibliography{sigproc} 


\begin{thebibliography}{00}


\ifx \showCODEN    \undefined \def \showCODEN     #1{\unskip}     \fi
\ifx \showDOI      \undefined \def \showDOI       #1{#1}\fi
\ifx \showISBNx    \undefined \def \showISBNx     #1{\unskip}     \fi
\ifx \showISBNxiii \undefined \def \showISBNxiii  #1{\unskip}     \fi
\ifx \showISSN     \undefined \def \showISSN      #1{\unskip}     \fi
\ifx \showLCCN     \undefined \def \showLCCN      #1{\unskip}     \fi
\ifx \shownote     \undefined \def \shownote      #1{#1}          \fi
\ifx \showarticletitle \undefined \def \showarticletitle #1{#1}   \fi
\ifx \showURL      \undefined \def \showURL       {\relax}        \fi
\providecommand\bibfield[2]{#2}
\providecommand\bibinfo[2]{#2}
\providecommand\natexlab[1]{#1}
\providecommand\showeprint[2][]{arXiv:#2}

\bibitem[\protect\citeauthoryear{Bovik}{Bovik}{2010}]%
        {bovik2010handbook}
\bibfield{author}{\bibinfo{person}{Alan~C Bovik}.}
  \bibinfo{year}{2010}\natexlab{}.
\newblock \bibinfo{booktitle}{{\em Handbook of image and video processing}}.
\newblock \bibinfo{publisher}{Academic press}.
\newblock


\bibitem[\protect\citeauthoryear{Brezeale and Cook}{Brezeale and Cook}{2008}]%
        {video_classifcation}
\bibfield{author}{\bibinfo{person}{Darin Brezeale} {and}
  \bibinfo{person}{Diane~J Cook}.} \bibinfo{year}{2008}\natexlab{}.
\newblock \showarticletitle{Automatic video classification: A survey of the
  literature}.
\newblock \bibinfo{journal}{{\em IEEE Transactions on Systems, Man, and
  Cybernetics, Part C (Applications and Reviews)\/}} \bibinfo{volume}{38},
  \bibinfo{number}{3} (\bibinfo{year}{2008}), \bibinfo{pages}{416--430}.
\newblock


\bibitem[\protect\citeauthoryear{Dewan, Rai~Dastidar, and Ahmad}{Dewan
  et~al\mbox{.}}{2018}]%
        {Dewan2018}
\bibfield{author}{\bibinfo{person}{Karan Dewan}, \bibinfo{person}{Tathagato
  Rai~Dastidar}, {and} \bibinfo{person}{Maroof Ahmad}.}
  \bibinfo{year}{2018}\natexlab{}.
\newblock \showarticletitle{Estimation of Sperm Concentration and Total
  Motility From Microscopic Videos of Human Semen Samples}. In
  \bibinfo{booktitle}{{\em Proceedings of the IEEE Conference on Computer
  Vision and Pattern Recognition (CVPR) Workshops}}.
\newblock


\bibitem[\protect\citeauthoryear{Farneb{\"a}ck}{Farneb{\"a}ck}{2003}]%
        {denseopticalflow}
\bibfield{author}{\bibinfo{person}{Gunnar Farneb{\"a}ck}.}
  \bibinfo{year}{2003}\natexlab{}.
\newblock \showarticletitle{Two-frame motion estimation based on polynomial
  expansion}. In \bibinfo{booktitle}{{\em Proceedings of the Scandinavian
  conference on Image analysis}}. Springer, \bibinfo{pages}{363--370}.
\newblock


\bibitem[\protect\citeauthoryear{Hampapur, Weymouth, and Jain}{Hampapur
  et~al\mbox{.}}{1994}]%
        {hampapur1994digital}
\bibfield{author}{\bibinfo{person}{Arun Hampapur}, \bibinfo{person}{Terry
  Weymouth}, {and} \bibinfo{person}{Ramesh Jain}.}
  \bibinfo{year}{1994}\natexlab{}.
\newblock \showarticletitle{Digital video segmentation}. In
  \bibinfo{booktitle}{{\em Proceedings of the second ACM international
  conference on Multimedia}}. ACM, \bibinfo{pages}{357--364}.
\newblock


\bibitem[\protect\citeauthoryear{Haugen, Hicks, Andersen, Witczak, Hammer,
  Borgli, Halvorsen, and Riegler}{Haugen et~al\mbox{.}}{2019}]%
        {visem}
\bibfield{author}{\bibinfo{person}{Trine~B. Haugen}, \bibinfo{person}{Steven~A.
  Hicks}, \bibinfo{person}{Jorunn~M. Andersen}, \bibinfo{person}{Oliwia
  Witczak}, \bibinfo{person}{Hugo~L. Hammer}, \bibinfo{person}{Rune Borgli},
  \bibinfo{person}{P{\aa}l Halvorsen}, {and} \bibinfo{person}{Michael~A.
  Riegler}.} \bibinfo{year}{2019}\natexlab{}.
\newblock \showarticletitle{VISEM: A Multimodal Video Dataset of Human
  Spermatozoa}. In \bibinfo{booktitle}{{\em Proceedings of the 10th ACM on
  Multimedia Systems Conference}} {\em (\bibinfo{series}{MMSys'19})}.
  \bibinfo{publisher}{ACM}, \bibinfo{address}{New York, NY, USA}.
\newblock
\showISBNx{78-1-4503-6297-9}
\showDOI{%
\url{https://doi.org/10.1145/3304109.3325814}}


\bibitem[\protect\citeauthoryear{He, Zhang, Ren, and Sun}{He
  et~al\mbox{.}}{2016}]%
        {resnet_34}
\bibfield{author}{\bibinfo{person}{Kaiming He}, \bibinfo{person}{Xiangyu
  Zhang}, \bibinfo{person}{Shaoqing Ren}, {and} \bibinfo{person}{Jian Sun}.}
  \bibinfo{year}{2016}\natexlab{}.
\newblock \showarticletitle{Deep residual learning for image recognition}. In
  \bibinfo{booktitle}{{\em Proceedings of the IEEE conference on computer
  vision and pattern recognition}}. \bibinfo{pages}{770--778}.
\newblock


\bibitem[\protect\citeauthoryear{Hicks, Halvorsen, Haugen, Andersen, Witczak,
  Pogorelov, Hammer, Dang-Nguyen, Lux, and Riegler}{Hicks
  et~al\mbox{.}}{2019}]%
        {medico2019overview}
\bibfield{author}{\bibinfo{person}{Steven Hicks}, \bibinfo{person}{P{\aa}l
  Halvorsen}, \bibinfo{person}{Trine~B Haugen}, \bibinfo{person}{Jorunn~M
  Andersen}, \bibinfo{person}{Oliwia Witczak}, \bibinfo{person}{Konstantin
  Pogorelov}, \bibinfo{person}{Hugo~L Hammer}, \bibinfo{person}{Duc-Tien
  Dang-Nguyen}, \bibinfo{person}{Mathias Lux}, {and} \bibinfo{person}{Michael
  Riegler}.} \bibinfo{year}{2019}\natexlab{}.
\newblock \showarticletitle{Medico Multimedia Task at MediaEval 2019}. In
  \bibinfo{booktitle}{{\em CEUR Workshop Proceedings - Multimedia Benchmark
  Workshop (MediaEval)}}.
\newblock


\bibitem[\protect\citeauthoryear{Itseez}{Itseez}{2014}]%
        {opencv}
Itseez \bibinfo{year}{2014}\natexlab{}.
\newblock \bibinfo{booktitle}{{\em The OpenCV Reference Manual\/}
  (\bibinfo{edition}{2.4.9.0} ed.)}.
\newblock Itseez.
\newblock


\bibitem[\protect\citeauthoryear{Karpathy, Toderici, Shetty, Leung, Sukthankar,
  and Fei-Fei}{Karpathy et~al\mbox{.}}{2014}]%
        {data_2014large}
\bibfield{author}{\bibinfo{person}{Andrej Karpathy}, \bibinfo{person}{George
  Toderici}, \bibinfo{person}{Sanketh Shetty}, \bibinfo{person}{Thomas Leung},
  \bibinfo{person}{Rahul Sukthankar}, {and} \bibinfo{person}{Li Fei-Fei}.}
  \bibinfo{year}{2014}\natexlab{}.
\newblock \showarticletitle{Large-scale video classification with convolutional
  neural networks}. In \bibinfo{booktitle}{{\em Proceedings of the IEEE
  conference on Computer Vision and Pattern Recognition}}.
  \bibinfo{pages}{1725--1732}.
\newblock


\bibitem[\protect\citeauthoryear{Kingma and Ba}{Kingma and Ba}{2014}]%
        {adam}
\bibfield{author}{\bibinfo{person}{Diederik~P Kingma} {and}
  \bibinfo{person}{Jimmy Ba}.} \bibinfo{year}{2014}\natexlab{}.
\newblock \showarticletitle{Adam: A method for stochastic optimization}.
\newblock \bibinfo{journal}{{\em arXiv preprint arXiv:1412.6980\/}}
  (\bibinfo{year}{2014}).
\newblock


\bibitem[\protect\citeauthoryear{Li, Min, Shen, Carlson, and Carin}{Li
  et~al\mbox{.}}{2018}]%
        {li2018video}
\bibfield{author}{\bibinfo{person}{Yitong Li}, \bibinfo{person}{Martin~Renqiang
  Min}, \bibinfo{person}{Dinghan Shen}, \bibinfo{person}{David Carlson}, {and}
  \bibinfo{person}{Lawrence Carin}.} \bibinfo{year}{2018}\natexlab{}.
\newblock \showarticletitle{Video generation from text}. In
  \bibinfo{booktitle}{{\em Thirty-Second AAAI Conference on Artificial
  Intelligence}}.
\newblock


\bibitem[\protect\citeauthoryear{Mortimer, van~der Horst, and
  Mortimer}{Mortimer et~al\mbox{.}}{2015}]%
        {mortimer2015future}
\bibfield{author}{\bibinfo{person}{Sharon~T Mortimer}, \bibinfo{person}{Gerhard
  van~der Horst}, {and} \bibinfo{person}{David Mortimer}.}
  \bibinfo{year}{2015}\natexlab{}.
\newblock \showarticletitle{The future of computer-aided sperm analysis}.
\newblock \bibinfo{journal}{{\em Asian journal of andrology\/}}
  \bibinfo{volume}{17}, \bibinfo{number}{4} (\bibinfo{year}{2015}),
  \bibinfo{pages}{545}.
\newblock


\bibitem[\protect\citeauthoryear{Pogorelov, Riegler, Halvorsen, Hicks, Randel,
  Dang-Nguyen, Lux, Ostroukhova, and de~Lange}{Pogorelov et~al\mbox{.}}{2018}]%
        {pogorelov2018medico}
\bibfield{author}{\bibinfo{person}{Konstantin Pogorelov},
  \bibinfo{person}{Michael Riegler}, \bibinfo{person}{P{\aa}l Halvorsen},
  \bibinfo{person}{Steven~Alexander Hicks}, \bibinfo{person}{Kristin~Ranheim
  Randel}, \bibinfo{person}{Duc-Tien Dang-Nguyen}, \bibinfo{person}{Mathias
  Lux}, \bibinfo{person}{Olga Ostroukhova}, {and} \bibinfo{person}{Thomas de
  Lange}.} \bibinfo{year}{2018}\natexlab{}.
\newblock \showarticletitle{Medico Multimedia Task at MediaEval 2018.}. In
  \bibinfo{booktitle}{{\em Proceedings of the CEUR Workshop on Multimedia
  Benchmark Workshop (MediaEval)}}.
\newblock


\bibitem[\protect\citeauthoryear{Riegler, Pogorelov, Halvorsen, Griwodz, Lange,
  Randel, Eskeland, Nguyen, Tien, Lux, et~al\mbox{.}}{Riegler
  et~al\mbox{.}}{2017}]%
        {riegler2017multimedia}
\bibfield{author}{\bibinfo{person}{Michael Riegler},
  \bibinfo{person}{Konstantin Pogorelov}, \bibinfo{person}{P{\aa}l Halvorsen},
  \bibinfo{person}{Carsten Griwodz}, \bibinfo{person}{Thomas Lange},
  \bibinfo{person}{Kristin Randel}, \bibinfo{person}{Sigrun Eskeland},
  \bibinfo{person}{Dang Nguyen}, \bibinfo{person}{Duc Tien},
  \bibinfo{person}{Mathias Lux}, {and} \bibinfo{person}{others}.}
  \bibinfo{year}{2017}\natexlab{}.
\newblock \showarticletitle{Multimedia for medicine: the medico task at
  MediaEval 2017}.
\newblock  (\bibinfo{year}{2017}).
\newblock


\bibitem[\protect\citeauthoryear{Soomro, Zamir, and Shah}{Soomro
  et~al\mbox{.}}{2012}]%
        {data_ucf101}
\bibfield{author}{\bibinfo{person}{Khurram Soomro},
  \bibinfo{person}{Amir~Roshan Zamir}, {and} \bibinfo{person}{Mubarak Shah}.}
  \bibinfo{year}{2012}\natexlab{}.
\newblock \showarticletitle{UCF101: A dataset of 101 human actions classes from
  videos in the wild}.
\newblock \bibinfo{journal}{{\em arXiv preprint arXiv:1212.0402\/}}
  (\bibinfo{year}{2012}).
\newblock


\bibitem[\protect\citeauthoryear{Srivastava, Hinton, Krizhevsky, Sutskever, and
  Salakhutdinov}{Srivastava et~al\mbox{.}}{2014}]%
        {dropout}
\bibfield{author}{\bibinfo{person}{Nitish Srivastava},
  \bibinfo{person}{Geoffrey Hinton}, \bibinfo{person}{Alex Krizhevsky},
  \bibinfo{person}{Ilya Sutskever}, {and} \bibinfo{person}{Ruslan
  Salakhutdinov}.} \bibinfo{year}{2014}\natexlab{}.
\newblock \showarticletitle{Dropout: A Simple Way to Prevent Neural Networks
  from Overfitting}.
\newblock \bibinfo{journal}{{\em Journal of Machine Learning Research\/}}
  \bibinfo{volume}{15} (\bibinfo{year}{2014}), \bibinfo{pages}{1929--1958}.
\newblock
\showURL{%
\url{http://jmlr.org/papers/v15/srivastava14a.html}}


\bibitem[\protect\citeauthoryear{Tulyakov, Liu, Yang, and Kautz}{Tulyakov
  et~al\mbox{.}}{2018}]%
        {tulyakov2018mocogan}
\bibfield{author}{\bibinfo{person}{Sergey Tulyakov}, \bibinfo{person}{Ming-Yu
  Liu}, \bibinfo{person}{Xiaodong Yang}, {and} \bibinfo{person}{Jan Kautz}.}
  \bibinfo{year}{2018}\natexlab{}.
\newblock \showarticletitle{Mocogan: Decomposing motion and content for video
  generation}. In \bibinfo{booktitle}{{\em Proceedings of the IEEE conference
  on computer vision and pattern recognition}}. \bibinfo{pages}{1526--1535}.
\newblock


\bibitem[\protect\citeauthoryear{{Urbano}, {Masson}, {VerMilyea}, and
  {Kam}}{{Urbano} et~al\mbox{.}}{2017}]%
        {Urbano2017}
\bibfield{author}{\bibinfo{person}{L.~F. {Urbano}}, \bibinfo{person}{P.
  {Masson}}, \bibinfo{person}{M. {VerMilyea}}, {and} \bibinfo{person}{M.
  {Kam}}.} \bibinfo{year}{2017}\natexlab{}.
\newblock \showarticletitle{Automatic Tracking and Motility Analysis of Human
  Sperm in Time-Lapse Images}.
\newblock \bibinfo{journal}{{\em IEEE Transactions on Medical Imaging\/}}
  \bibinfo{volume}{36}, \bibinfo{number}{3} (\bibinfo{date}{March}
  \bibinfo{year}{2017}), \bibinfo{pages}{792--801}.
\newblock
\showISSN{0278-0062}
\showDOI{%
\url{https://doi.org/10.1109/TMI.2016.2630720}}


\bibitem[\protect\citeauthoryear{Yue-Hei~Ng, Hausknecht, Vijayanarasimhan,
  Vinyals, Monga, and Toderici}{Yue-Hei~Ng et~al\mbox{.}}{2015}]%
        {video_classfication}
\bibfield{author}{\bibinfo{person}{Joe Yue-Hei~Ng}, \bibinfo{person}{Matthew
  Hausknecht}, \bibinfo{person}{Sudheendra Vijayanarasimhan},
  \bibinfo{person}{Oriol Vinyals}, \bibinfo{person}{Rajat Monga}, {and}
  \bibinfo{person}{George Toderici}.} \bibinfo{year}{2015}\natexlab{}.
\newblock \showarticletitle{Beyond short snippets: Deep networks for video
  classification}. In \bibinfo{booktitle}{{\em Proceedings of the IEEE
  conference on computer vision and pattern recognition}}.
  \bibinfo{pages}{4694--4702}.
\newblock


\end{thebibliography}

\end{document}